# Pressure enhanced superconductivity at 10 K in La doped EuBiS$_2$F

Gohil S. Thakur[1], Rajveer Jha[2], Zeba Haque[1], V.P.S. Awana[2], L. C. Gupta[1][†] and A. K. Ganguli[1,3][(a)]

[1]*Department of Chemistry, Indian Institute of Technology, New Delhi, India, 110016*

[2]*Quantum Phenomena and Applications Division, National Physical Laboratory (CSIR), New Delhi 110012, India*

[3]*Institute of Nano Science and Technology, Mohali, Punjab, India, 160064*

**Abstract**

We have investigated the effect of La doping and high pressure on superconducting properties of EuBiS$_2$F which is a newly discovered superconducting material ($T_c \sim 0.3$ K) [*Phys. Rev. B* **90**, 064518 (2014)]. An enhancement of $T_c$ to 2.2 K is observed in Eu$_{0.5}$La$_{0.5}$BiS$_2$F. Upon application of pressure $T_c$ is further enhanced upto ~10 K (P = 2.5 GPa). Eu$_{0.5}$La$_{0.5}$BiS$_2$F is semiconducting down to 3 K. An onset of superconductivity like feature is seen at 2.2 K at ambient pressure. At a pressure above 1.38GPa, the $T_c^{onset}$ remains invariant at 10 K but the $T_c(\rho=0)$ is increased to above 8.2 K. There is a possible crystallographic transformation by application of pressure from a structure of low $T_c$ to a structure corresponding to high $T_c$.

## 1. Introduction

Tremendous surge for finding new materials or modifying the existing materials for various applications has led to discovery of superconductivity in many compounds which were previously known but not investigated for their superconducting properties. LnOBiS$_2$ is one such family of materials. Discovery of superconductivity in Bi$_4$O$_4$S$_3$ at 4.5 K kick-started the research in layered Bi−S based materials [1,2]. This led to discovery of superconductivity in F-doped LnOBiS$_2$ (Ln = La, Ce, Pr, Nd and Yb; with $T_c$ ~ 2.5 − 10 K) [3–10]. Undoped LnOBiS$_2$ is a band semiconductor/insulator [11]. Structurally it consists of alternating rock-salt type BiS$_2$ layers and fluorite type edge-shared tetrahedral Ln$_2$O$_2$ layers [12]. Superconductivity could be induced in these materials by electron doping; either by F- substitution at O sites or by tetravalent ion (Ti, Zr, Hf and Th) substitution at Ln sites [3,4,8–10]. Hole doping has been unsuccessful in inducing superconductivity so far [13]. $T_c$ is low (2.5−5 K) [9] for the electron-doped samples but could be enhanced upto 10 K either by annealing samples under high pressure [3,14,15] or by application of external high pressure [16–19]. Very recently a new structurally analogous compound SrBiS$_2$F [20] was reported which itself is an semiconductor even under high pressure [21]. It becomes superconducting only upon doping with electrons (via La or Ce substitution at Sr site) [22,23]. It is noteworthy that $T_c$ increases significantly upon applying hydrostatic pressure in this material [24,25]. Since Eu$^{+2}$ and Sr$^{+2}$ are similar in size, hence replacement of Sr by Eu to form EuBiS$_2$F was tried successfully by *Zhai et al* [26]. However unlike SrBiS$_2$F, EuBiS$_2$F compound was surprisingly found to be metallic and even superconducting though with very low $T_c$ (~ 0.3 K). A mixed valence of Eu present in this material leads to self doping and superconductivity [26]. Till now there has been no report of any kind of doping or high pressure studies on EuBiS$_2$F. In this short manuscript we report enhancement of $T_c$ to 2.2 K in $Eu_{0.5}La_{0.5}BiS_2F$ and further enhancement of $T_c$ to an impressive 10 K by applying external pressure. To our knowledge, this is the first case where such a drastic enhancement in $T_c$ has been observed in Bi-S based materials.

**2. Experimental –** Polycrystalline sample of nominal composition $Eu_{0.5}La_{0.5}BiS_2F$ was synthesized via solid state method by the reaction of $EuF_3$, $La_2S_3$, $Bi_2S_3$ and Eu metal. $La_2S_3$ and $Bi_2S_3$ were presynthesized by the reaction of their respective elements at 800°C and 500°C respectively for 15 h in vacuum. Stoichiometric amounts of reactants were mixed well and sealed in an evacuated quartz tube and heated at 800°C for 24 hours. The resultant product was reground, pelletized and heated again under same conditions. The pellet was black, without luster and stable in air for weeks. Sample was characterized by powder x-ray diffraction technique using Cu−K$\alpha$ radiation ($\lambda$ = 1.5418 Å). For crystal structure analysis, Rietveld refinement of the powder XRD data was performed by using *Topas* software. Pressure dependent resistivity measurements were carried out by Physical Property Measurements System (*PPMS*-14T, *Quantum Design*). Pressure was applied using HPC−33 Piston type pressure cell with Quantum design DC resistivity option. Hydrostatic pressures were generated by a BeCu/NiCrAl clamped piston-cylinder cell. The sample was immersed in a fluid (Daphne Oil) pressure transmitting medium in a Teflon cell. Annealed Pt wires were affixed to gold-sputtered contact surfaces on each sample with silver epoxy in a standard four-wire configuration.

**3. Results and discussion –** Powder X-ray diffraction pattern of the sample confirms the presence of $EuBiS_2F$ phase as shown in figure 1. Some minor impurity peaks were also detected which correspond to $Bi_2S_3$ (*) and (#) $EuF_{2.4}$ (<10 % total). The sample crystallizes in tetragonal $CeOBiS_2$ structure [12] with P4/*nmm* space group (#129). Crystal structure of (Eu,La)BiS$_2$F is presented in figure 1 (b). It consists of Eu$_2$F$_2$ edge shared tetrahedral layers alternating with square pyramidal BiS$_2$ layers. Structural parameters were calculated by Rietveld fitting of powder X-ray diffraction data in the 2θ range 10 – 70°, results of which are shown in Table 1. The obtained lattice parameters were *a* = 4.0744(1) Å and *c* = 13.3336(7) Å and volume *V* = 221.35(2) Å$^3$. The *c*-parameter and unit cell volume for La-doped EuBiS$_2$F are markedly smaller than those of the undoped EuBiS$_2$F (*c* =13.5338 Å and *V* = 222.076 Å$^3$) [26]. This reduction in the cell parameter is indicative of effective doping as La$^{3+}$ ion (1.16 Å) is smaller as compared to Eu$^{2+}$ (1.25 Å) assuming VIII coordination.

Resistivity measurements at ambient pressure for Eu$_{0.5}$La$_{0.5}$BiS$_2$F are shown in figure 2 (a). A small drop in resistivity observed at 2.2 K indicates an onset of superconductivity. We do not observe zero resistance at ambient pressure as our measurements were limited down to 2 K only. Certain features of these resistivity measurements deserve to be pointed out. 1) An anomaly in resistivity around 280 K associated with a charge density wave in undoped EuBiS$_2$F. This has been completely suppressed. 2) A semiconducting like normal state behavior is observed (inset of figure 2 (a)) in our sample which is in contrast with the metal like behavior observed in undoped material below CDW (280 K)[26]. This kind of semiconducting behavior is also observed in LnO$_{1-x}$F$_x$BiS$_2$ [3,4,9] and Sr$_{0.5}$Re$_{0.5}$FBiS$_2$ [22,23] at ambient pressure. Application of pressure creates marked effects on these features. Resistivity as a function of temperature under variable pressure is presented in figure 2 (b). As the applied pressure increases the semiconducting behavior is gradually suppressed and the superconducting transition becomes more and more clearly visible. A gradual decreasing trend in resistivity is observed with pressure. The drop in resistivity at 2.2 K (at P = 0 GPa) indicating the $T_c$ onset gradually shifts to higher temperatures with application of pressure. This drop in resistivity seems pronounced at higher pressure (upto P = 0.55 GPa). Also at this pressure a kink in resistivity starts appearing at ~ 9 K. This kink could be attributed to superconductivity in a high $T_c$ phase as it becomes prominent with increasing the pressure (see figure 3). It is to be note that a zero resistance state is not observed till a pressure of 0.55 GPa and only at 0.97GPa did we observe a zero resistance state with a broad transition width ($T_c^{onset}$ of 10 K and $T_c^{zero}$ ~ 3 K (figure 3)). On increasing the pressure beyond 1 GPa, the transition width decreases and a clear superconducting state is achieved with $T_c$ onset remaining unchanged ~ 10 K but $T_c^{zero}$ shifts to higher temperature with a maximum of 8.2 K for pressure ~ 2.5 GPa (figure 3). The criterion on selection of $T_c$ is demonstrated in figure 3. The semiconducting state is suppressed completely and a metallic behavior is observed in the normal state at high pressures. Such change of semiconducting to metallic state coupled with an appearance of high temperature superconducting state is also observed in $Sr_{0.5}Re_{0.5}FBiS_2$ under pressure [24,25]. The sudden jump in $T_c$ near 1 GPa indicates a possible transformation of a low $T_c$ phase ($T_c$ < 4 K) to a high $T_c$ phase ($T_c$ ~ 9−10 K). This jump is clearly evident from $T_c$ v/s pressure phase diagram as shown in figure 4. There are already few examples reported in literature which show such kind of phase transformation on application of pressure [18,19,27]. But this is the first time it is reported for $EuBiS_2F$. A possible structural transformation to a lower symmetry is not ruled out as observed in $La(O,F)BiS_2$ [18] and very recently in $EuFBiS_2$[28,29].

Moreover to estimate the upper critical field, $H_{c2}$ magnetic field dependent resistivity at constant pressure (2.5GPa) has been measured and results are shown in figure 5(a). $T_c$ onset shifts only slightly upon application of magnetic field but $T_c^{zero}$ decreases very readily. The value of $H_{c2}(0)$ is estimated using WHH formula; $H_{c2}(0) = -0.693T_c (dH_{c2}/dT)$ [30]. The $H_{c2}(T)$ curve fitted with WHH equation is shown in figure 5b which gives the estimated value of $H_{c2}(0) = 10$ T. This value of $H_{c2}$ is significantly smaller than those obtained for $Sr_{0.5}Ln_{0.5}FBiS_2$ under pressure (14 − 20 T) [24,25]. The upper critical field as estimated by WHH model is quite smaller than the Pauli paramagnetic limit $H_p = 1.84T_c = 18$ T. We also observe flattening of $H_{c2}$ curve at high fields (~5 T) which we think is due to Pauli spin paramagnetic effect.

**4. Conclusions –** We have observed superconductivity with $T_c$ (onset) as high as 10 K in $Eu_{0.5}La_{0.5}BiS_2F$ under pressure. The possible charge density wave occurring at 280 K in undoped material is suppressed upon electron doping and a semiconducting behavior is observed before the onset of superconductivity at 2.2 K. Application of external pressure further increases the $T_c$ upto an impressive value of 10 K. This is the first report of significant enhancement of $T_c$ by electron doping and high pressure in $EuBiS_2F$. We also speculate a possible phase transformation on application of pressure.

**Acknowledgments**

GST and ZH thanks CSIR and UGC, India, respectively for a fellowship. This work is partially supported financially by DST, Government of India (AKG) and *DAE-SRC* outstanding investigator award scheme on search for new superconductors at NPL-CSIR, India.

**Note:** [†]Visiting scientist at Solid State and Nano Research Laboratory, Department of chemistry, IIT Delhi, India.

**Figure 1. (a)** Rietveld fitted room temperature powder X-ray diffraction data of $Eu_{0.5}La_{0.5}BiS_2F$ and **(b)** crystal structure of $Eu_{0.5}La_{0.5}BiS_2F$ (right). (*) indicates the $Bi_2S_3$ impurity phase.

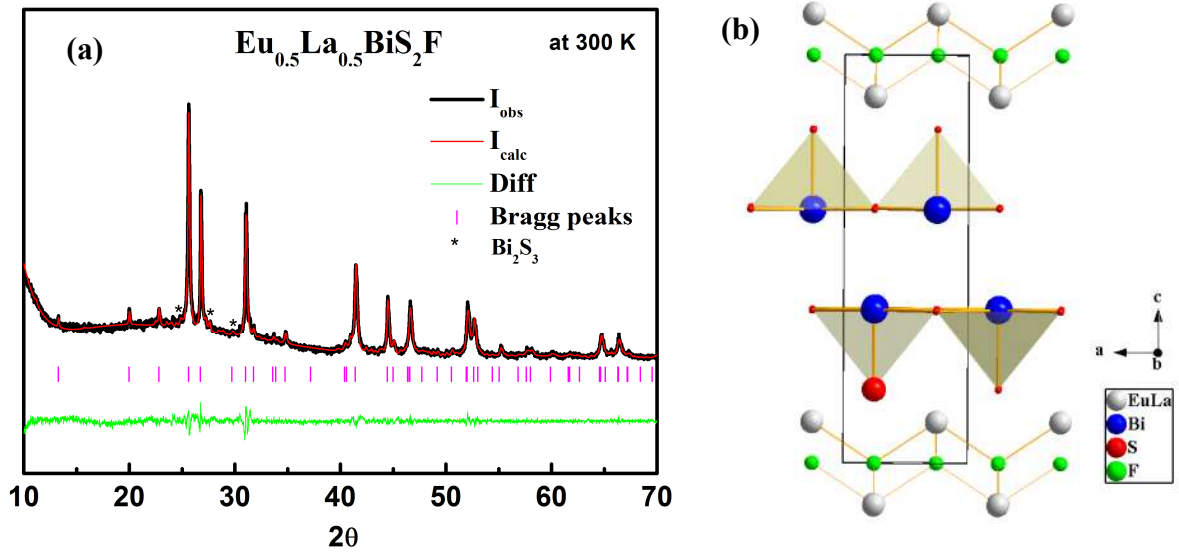

**Table 1.** Refined structural parameters for $Eu_{0.5}La_{0.5}BiS_2F$.

| Space Group | P4/nmm |
|---|---|
| $R_{wp}$ (%) | 5.32 |
| $\chi^2$ | 2.54 |
| $a = b$ (Å) | 4.0744(1) |
| $c$ (Å) | 13.3336(7) |
| $V$ (Å$^3$) | 221.35(2) |

| Atom | site | x | y | z | Occu (fixed) |
|---|---|---|---|---|---|
| Eu/La | 2c | 0.25 | 0.25 | 0.119(3) | 0.5/0.5 |
| F | 2a | 0.75 | 0.25 | 0 | 1 |
| Bi | 2c | 0.25 | 0.25 | 0.629(2) | 1 |
| S1 | 2c | 0.25 | 0.25 | 0.397(7) | 1 |
| S2 | 2c | 0.25 | 0.25 | 0.835(5) | 1 |

**Figure 2. (a)** Variable temperature resistivity plot for $Eu_{0.5}La_{0.5}BiS_2F$ at ambient pressure in the low temperature range. Inset show the plot of resistivity in full temperature range upto 300 K. **(b)** Resistivity plots for $Eu_{0.5}La_{0.5}BiS_2F$ at various applied pressures.

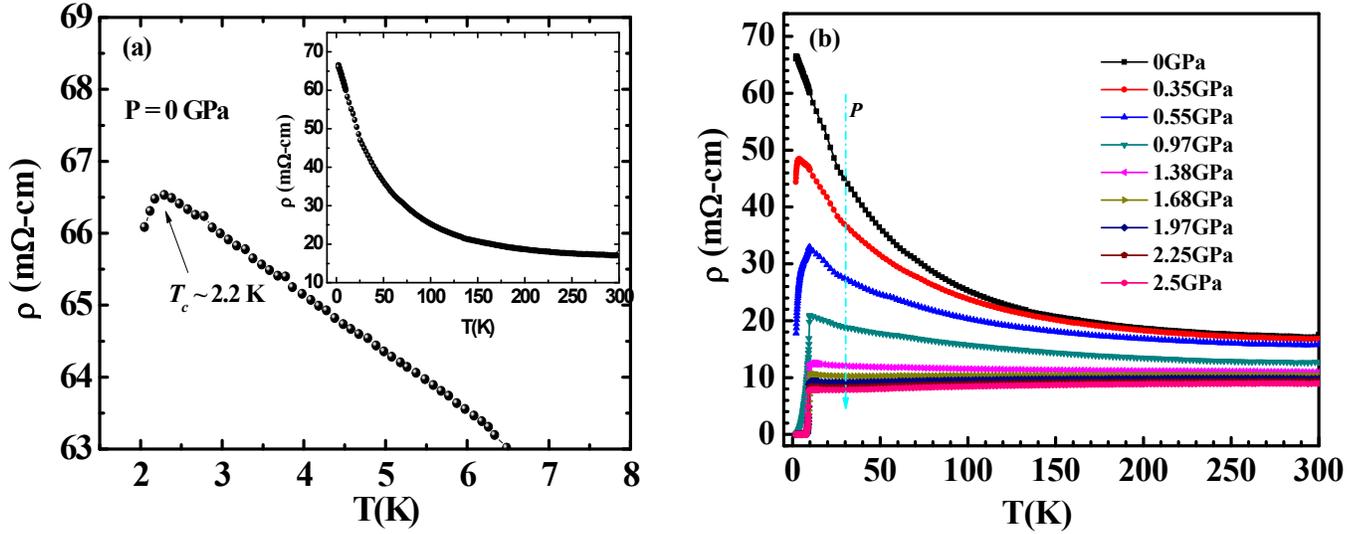

**Figure 3.** Magnified resistivity plots for $Eu_{0.5}La_{0.5}BiS_2F$ at different applied pressures. The violet lines show the criterion on selection of $T_c$ values. Arrow marks the origin of kink in resistivity.

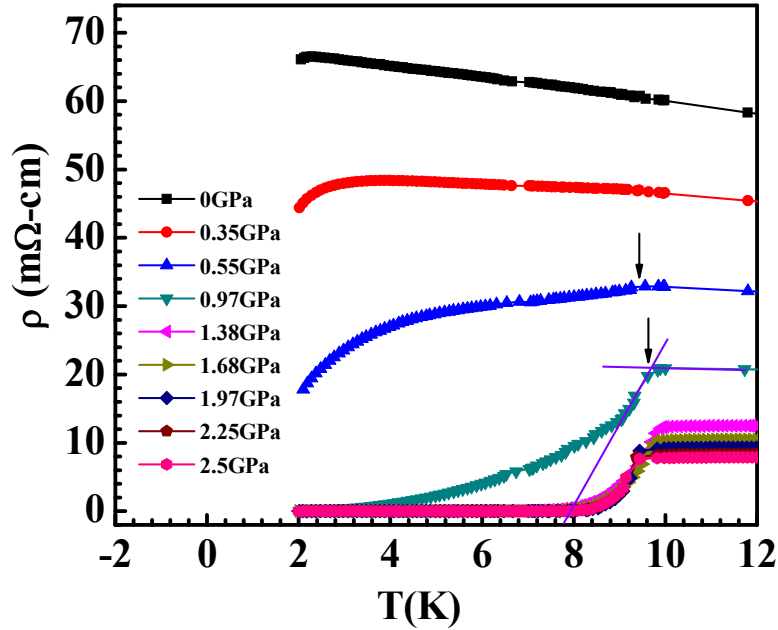

**Figure 4.** Pressure vs. $T_c$ phase diagram for $Eu_{0.5}La_{0.5}BiS_2F$ showing regions of low and high $T_c$.

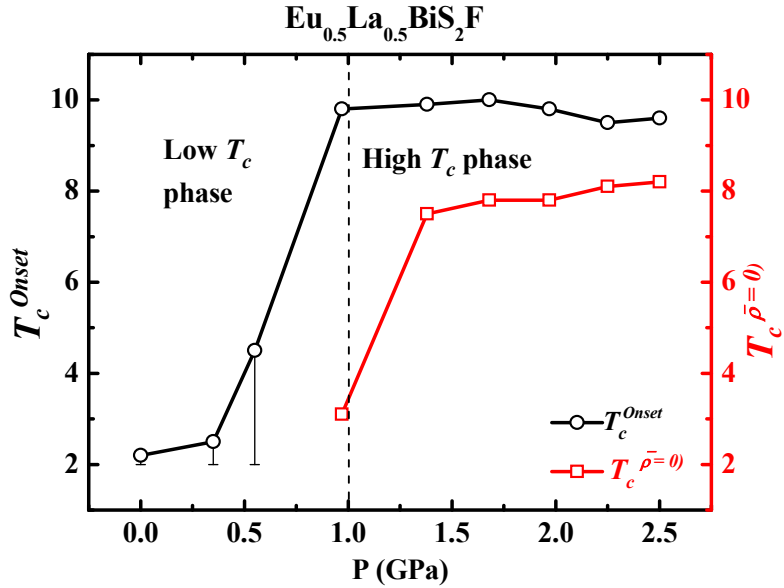

**Figure 5 (a)** Magnetic field dependent resistivity plots for $Eu_{0.5}La_{0.5}BiS_2F$ at an applied pressure of 2.5 GPa. **(b)** $H_{c2}(T)$ curves for $Eu_{0.5}La_{0.5}BiS_2F$ at P = 2.5 GPa; black squares are the experimental data points and red line is the WHH fit.

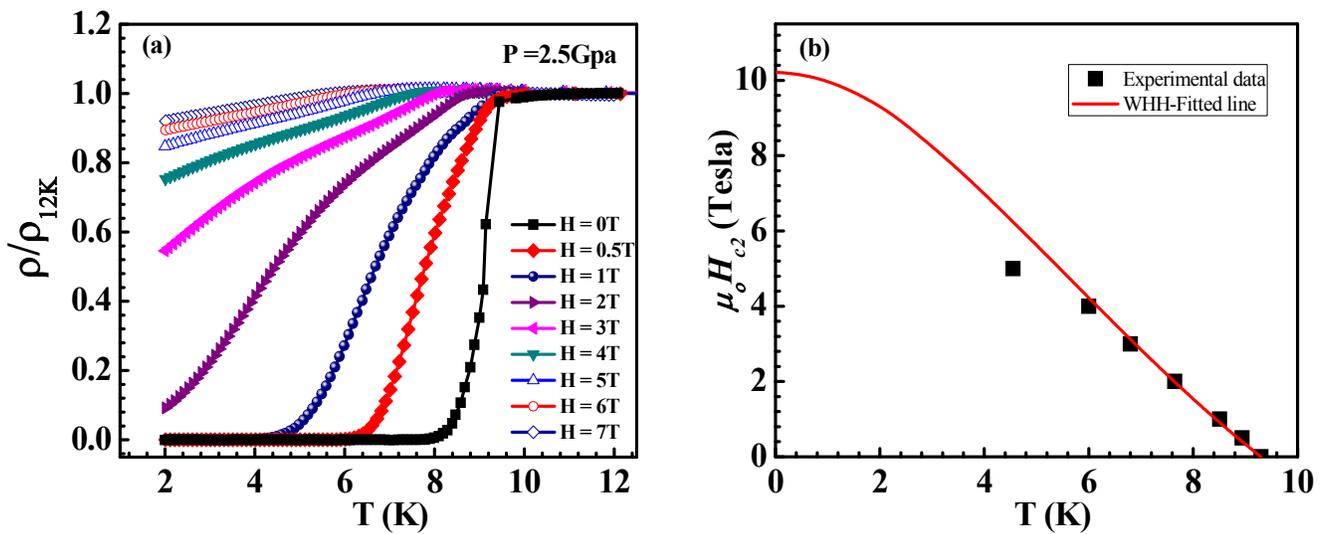